\documentclass[sigconf,authorversion,nonacm]{acmart}

\AtBeginDocument{%
  }

\setcopyright{acmcopyright}
\copyrightyear{2023}
\acmYear{2023}
\acmDOI{XXXXXXX.XXXXXXX}

\acmConference[]{Second International workshop on RISC-V for HPC}{November 13,
  2023}{Denver, CO}
\acmPrice{15.00}
\acmISBN{978-1-4503-XXXX-X/18/06}

\usepackage{tikz}
\usepackage{pgfplots}
\pgfplotsset{compat=1.18} 
\usepackage{xfrac}

\usepackage{subcaption}
\usepackage{listings}%

\newcommand{\cpp}[1]{\lstinline[language=c++,keywordstyle ={\color{azure}},morekeywords={co_return,co_yield,co_await,ensure_started,sync_wait,start_detached,any_sender}]{#1}}

\usepackage{xcolor} 
\definecolor{azure}{rgb}{0.0, 0.5, 1.0}
\definecolor{darkgreen}{rgb}{0.0, 0.5, 0.0}
\definecolor{amaranth}{rgb}{0.9, 0.17, 0.31}
\definecolor{cadetgrey}{rgb}{0.57, 0.64, 0.69}
\definecolor{aureolin}{rgb}{0.99, 0.93, 0.0}

\begin{document}

\title{Evaluating HPX and Kokkos on RISC-V using an Astrophysics Application Octo-Tiger}

\author{Patrick Diehl}
\orcid{0000-0003-3922-8419}
\author{Gregor Dai\ss{}}
\author{Steven R. Brandt}
\author{Alireza Kheirkhahan}
\author{Hartmut Kaiser}
\orcid{0000-0002-8712-2806}
\email{{pdiehl,sbrandt,aliz,hkaiser}@cct.lsu.edu,gdaiss1@lsu.edu}

\affiliation{%
  \institution{Center of Computation \& Technology \\ Louisiana State University}
  \streetaddress{Digital Media Center}
  \city{Baton Rouge}
  \state{Louisiana}
  \country{USA}
  \postcode{70803}
}

\author{Christopher Taylor}
\author{John Leidel}
\affiliation{%
  \institution{Tactical Computing Labs}
  \streetaddress{1001 Pecan St.}
  \city{Lindsay, Texas}
  \country{USA}}
\email{{ctaylor,jleidel}@tactcomplabs.com}

\renewcommand{\shortauthors}{Diehl et al.}

\begin{abstract}
In recent years, computers based on the RISC-V architecture have raised broad interest in the high-performance computing (HPC) community. As the RISC-V community develops the core instruction set architecture (ISA) along with ISA extensions, the HPC community has been actively ensuring HPC applications and environments~\cite{riscv_sig,riscv_shah} are supported. In this context, assessing the performance of asynchronous many-task runtime systems (AMT) is essential. In this paper, we describe our experience with porting of a full 3D adaptive mesh-refinement, multi-scale, multi-model, and multi-physics application, Octo-Tiger, that is based on the HPX AMT, and we explore its performance characteristics on different RISC-V systems. Despite the (limited) capabilities of the RISC-V test systems we used, Octo-Tiger already shows promising results and good scaling. We, however, expect that additional hardware support based on dedicated ISA extensions (such as single-cycle context switches, extended atomic operations, and direct support for HPX's global address space) would allow for even better performance results.
\end{abstract}




\maketitle

\section{Introduction}
\textit{RISC-V} was introduced to the open community in 2015 as an open standard instruction set architecture (ISA) which is an iteration on established reduced instruction set computer (RISC) principles~\cite{waterman2014risc}. RISC-V is based on long academic and practical experience, and an international standards committee with broad industry support maintains its design. Currently, most general-purpose CPUs are based on proprietary ISAs, \emph{e.g.}\ Intel and AMD use the x86 ISA. Riken's Supercomputer\ Fugaku recently used A64FX CPU based on the Arm ISA. However, none of these ISAs can be used without permission and payment of royalties. RISC-V, on the contrary, is completely open for use by anyone and royalty-free. For these and other reasons, the European Processor Initiative (EPI), which aims to develop a vendor-independent European CPU for high-performance computing, has identified RISC-V as a target for future investment.

This paper explores the porting and performance of the C\texttt{++} standard library for parallelism and concurrency (HPX)~\cite{kaiser2020hpx,kaiser2023hpx} to a RISC-V single-board computer (SBC). HPX is an asynchronous many-task runtime system (AMT) targeting any-scale from single board computers (like Raspberry Pi~\cite{gupta2020deploying}) to supercomputers (like ORNL's Summit which is based on IBM\textsuperscript{\textregistered}'s Power\ 9\texttrademark~CPUS~\cite{diehl2021octo}, CSCS's Piz\ Daint using Intel\textsuperscript{\textregistered} x86 CPUs~\cite{10.1145/3295500.3356221}, or Riken's Supercomputer\ Fugaku using Fujitsu\textsuperscript{\textregistered} A64FX CPUs~\cite{diehl2023simulating}).

The three RISC-V systems mentioned in this paper use development boards and do not have HPC-grade hardware components. This paper uses a SiFive HiFive Unmatched development board with four physical cores, a Pine64 Star64 RISC-V board with a four physical core StarFive processor (a licensed SiFive JH7110 design), and two VisionFive2 Open Source RISC-V boards with four physical cores for the distributed runs. At the time of this writing, no HPC-grade RISC-V hardware was available. However, in this paper, we already prepare for the release of such HPC RISC-V hardware by both porting and our HPC astrophysics application Octo-Tiger and its toolchain (notably HPX) to RISC-V and evaluating their performance on the currently available hardware. In addition, we will investigate the performance of a simple benchmark on RISC-V with traditional HPC CPUs, like Intel, AMD, and Arm A64FX.

The paper is structured as follows: Section~\ref{sec:related:work} discusses the related work. The software stack is introduced in Section~\ref{sec:software:stack}. Our in-house RISC-V cluster is presented in Section~\ref{sec:in:house:cluster}. Section~\ref{sec:porting:software:stack} describes the effort to port the software stack to RISC-V. Performance measurements are shown in Section~\ref{sec:performance}. The energy consumption is discussed in Section~\ref{sec:energy}. Section~\ref{sec:conclusion} finally concludes the paper.

\section{Related Work}
\label{sec:related:work}
The RISC-V architecture has been explored for the following applications: cryptography~\cite{stoffelen2019efficient}, deep learning~\cite{louis2019towards}, and internet of things~\cite{schiavone2017slow}. Different RISC-V CPUs from various vendors are available, and a comparison of selected CPUs is available here~\cite{10.1145/3457388.3458657}. As of this writing, RISC-V has not yet been widely adopted in scientific computing or high performance computing, but future adoption looks promising.

It sometimes takes time for an ISA to gain traction. For comparison, the ARM64 (Armv8-Series) architecture with 64 Bit was released in 2013, but it took until May 2020 for it to be used in Riken's Supercomputer\ Fugaku Top 500 machine. The RISC-V ISA was released in 2015, but its adoption in the HPC community has been slower than ARM64's.


This paper will discuss the porting of the C\texttt{++} standard library for parallelism and concurrency (HPX) to RISC-V. Since our performance section focuses on distributed runs on two single-board computers, we will also focus on the related work on distributed asynchronous many-task runtime systems. Notable mentions are Charm++~\cite{kale1993charm++}, Chapel~\cite{chamberlain2007parallel}, Legion~\cite{bauer2012legion}, and PaRSEC~\cite{bosilca2013parsec}. For a detailed comparison, we refer to~\cite{thoman2018taxonomy}. From the programming paradigm, HPX and Charm\texttt{++} support similar features. However, HPX's API is based on the specification within the C\texttt{++} standard, and Charm\texttt{++} is a library implemented in C\texttt{++}. A comparison between Charm\texttt{++} and HPX is available here~\cite{10.1007/978-3-031-31209-0_1}.

\section{Software stack}
\label{sec:software:stack}
In this work, we use two benchmarks. First, a simple HPX benchmark, implementing the Maclaurin series---second, a real-world HPX application: Octo-Tiger.

\begin{figure}[tb]
    \centering
    \includegraphics[width=\linewidth,trim={0 1cm 0 1cm},clip]{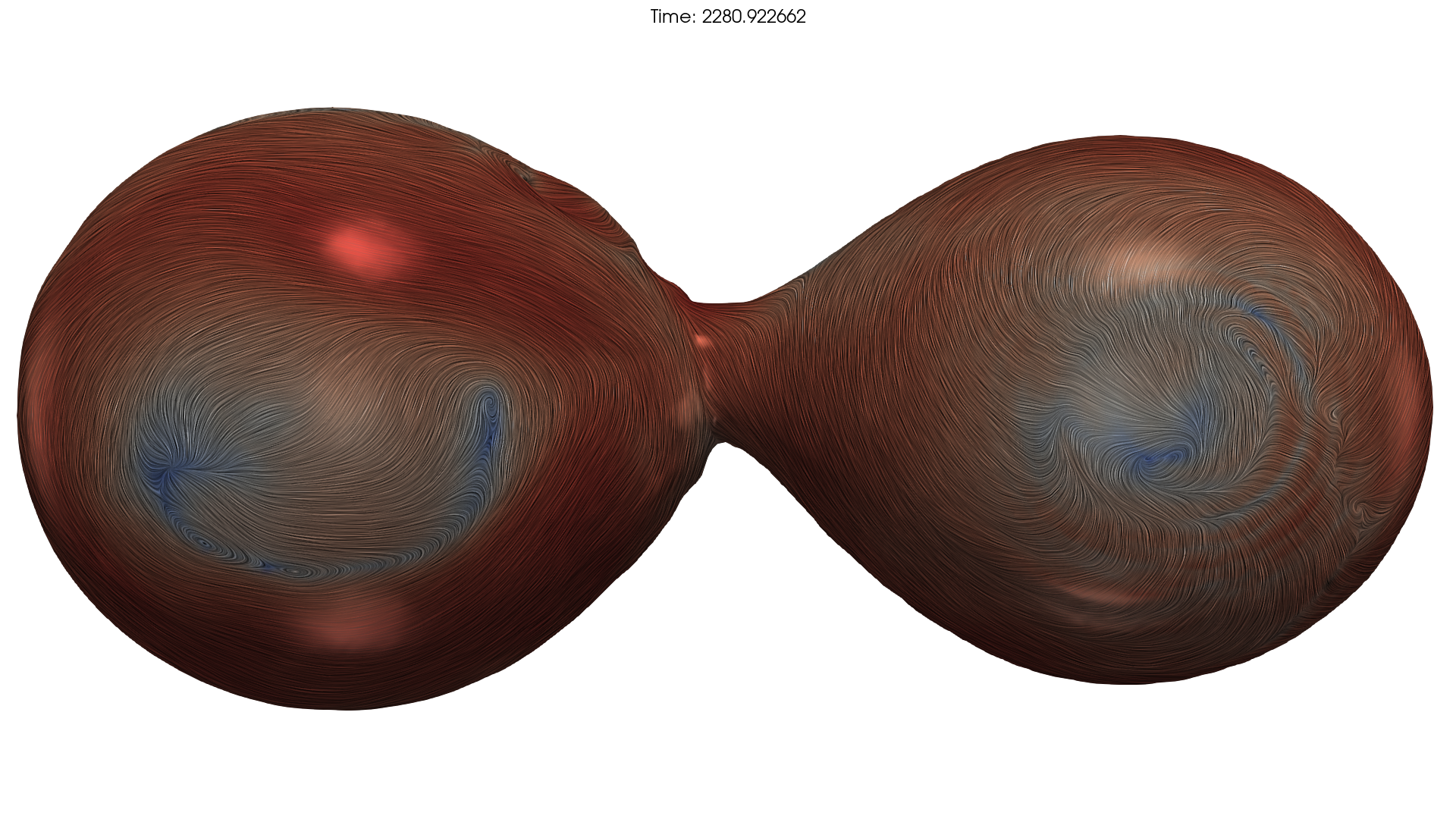}
    \caption{Merger of two stars with the aggregation belt between the two stars. The mass from the donor is transferred to the larger star. The color shows the velocity
magnitude of the stream, with red being high velocities. Adapted from~\cite{10.1145/3585341.3585354}.}
    \label{fig:enter-label}
\end{figure}

Octo-Tiger is a distributed astrophysical simulation based on HPX to simulate binary star systems (see Figure~\ref{fig:enter-label}).
It features an adaptive, tree-based data-structure,  interleaved solvers, and distributed and GPU capabilities. These features and the simulation's demanding compute requirements make Octo-Tiger an exciting benchmark for HPC work.

To provide context and illustrate the application's complexity that we ported to RISC-V, we briefly introduce Octo-Tiger and its main dependencies (HPX and Kokkos) in this section. We specifically focus on how and why each dependency is used in Octo-Tiger.
\subsection{HPX}
HPX is an Asynchronous Many-Task Runtime System (AMT), written in modern C\texttt{++}~\cite{kaiser2020hpx,kaiser_hartmut_2023_8216176}.
HPX makes it possible to write High Performance Computing (HPC) codes such as Octo-Tiger in a task-based fashion using C\texttt{++} futures and continuations. This yields a user-defined task graph. The parallelism within the graph is automatically used to hide communication latencies with remote nodes or synchronizations of GPU results.

HPX offers various node-level and distributed features to accelerate parallel codes.
At the node-level, HPX offers lightweight HPX threads (tasks) which can easily be suspended without blocking any OS threads. HPX allows tasks to be chained together using continuations (\lstinline{hpx::future::then, hpx::when_all}). In this way, a user can build a directed acyclic task graph (DAG) of tasks directly in their source code. Each of these asynchronous functions can return futures, making it possible to asynchronously wait for GPU work or communication. Furthermore, HPX implements the C\texttt{++} 20 API related to concurrency and parallelism. For instance, there is a \lstinline{hpx::mutex} one can use instead of the \lstinline{std::mutex} equivalent, see:~\cite{cxx17_standard,cxx20_standard}. The advantage to the HPX mutex is that the runtime can switch it out instead of simply blocking, allowing worker threads to continue working.

In Octo-Tiger, we use these node-level features to orchestrate kernel launches, data-transfers, and pre- and post-processing tasks for each time-step. This task-based approach is particulary handy for quickly making parallel work available to the system during the tree-traversals.

As mentioned, HPX also includes various distributed features. It includes an Active Global Address Space (AGAS \cite{9460481}), allowing HPX components to be distributed across multiple compute nodes. Remote function calls for methods of these components are available (with unified syntax between local and remote function calls). Furthermore, there are various useful communication primitives, such as buffers and channels, for data exchange.

Users can choose between the backend (parcelport) HPX uses for communication. Currently, three variants are available: TCP, MPI and LCI.

In Octo-Tiger, we use these features to distribute data across compute nodes (by having one HPX component per tree-node, which can be placed on any compute node available). Thanks to the unified syntax between local and remote function calls, the implementation of distributed tree traversals is greatly simplified, as we do not need to worry about, for example, if a child tree-node is on the same compute node when traversing the tree via recursion. The HPX futures we get from these asynchronous, potentially remote, function calls, and the futures obtained from data exchanges with other compute nodes neatly integrate with the aforementioned node-level task-graph, yielding a natural overlapping of computation and communication when enough tasks are available.

\subsection{Using Kokkos in HPX Applications}
HPX orchestrates the launches of the various compute kernels and manages their launch and data dependencies (by allowing the user to define them in the task-graph with futures), but what about the kernels themselves?
Given the heterogeneity of the currently available supercomputers, ranging from CPU-only platforms with AVX or SVE SIMD instructions to GPU supercomputers with Intel, AMD, or NVIDIA GPUs, writing portable compute kernels is crucial.

The Kokkos framework offers a programming model for developing such portable kernels~\cite{9485033}.
Kokkos provides various abstractions, like execution and memory spaces for various backends targeting both CPUs and GPUs of different vendors, helping developers achieve this desired (performance) portability.
Compute kernels written with Kokkos, using Kokkos Views as data-structures, can be adapted for the desired device by compiling them with the appropriate execution and memory spaces.
Thus, by simply using the correct memory spaces at compile-time, the same Kokkos compute kernel implementation can be run on NVIDIA GPUs (using Kokkos' CUDA execution space), on AMD GPUs (using the HIP execution space), and on Intel GPUs (using the SYCL execution space).

Of course, Kokkos kernels also work on CPUs. For example, by using a simple serial execution space (with one CPU core simply executing the kernel), or by using a parallel OpenMP execution space.
For better performance on CPUs, SIMD types are available: Operations on those compile down to the appropriate SIMD instructions (for example, AX512) when the Kokkos kernel is built for a CPU execution space. Alternatively, they are mapped to scalar operations on GPUs for compatibility, allowing us to run the same kernel with explicit SIMD vectorization on the CPU~\cite{sahasrabudhe2020portable} and scalar instructions on the GPU.

Kokkos works well with HPX due to two integrations. First, it is possible to get HPX futures for asynchronously launched Kokkos kernels, even on GPUs (which required event polling and integration with the underlying APIs such as CUDA \cite{daiss2021beyond} and SYCL~\cite{10.1145/3585341.3585354}).

The second integration is more critical for CPUs. There is a Kokkos HPX execution space that runs a Kokkos kernel directly on the HPX worker threads by internally splitting it into HPX tasks. This avoids conflicting thread pools (as we would encounter when trying to use the OpenMP execution space in HPX applications) and provides users with fine-grained control regarding the number of tasks that are required for each kernel (thus also steering the maximum number of cores desired per kernel which is helpful for concurrent kernels). Full system utilization can be achieved by running one large kernel divided into enough tasks for the available cores or by running enough small kernels concurrently, even if each uses one task (and thus one core). This is an ideal fit for an application such as Octo-Tiger with its adaptive data-structure and interleaved solvers as it contains a mix of computing tasks of varying intensity on each node.

We recently ported Octo-Tiger to Kokkos and now offer a Kokkos implementation for all major compute kernels in its solvers. Notably, both the Serial and the HPX execution space work for us when running the kernels on a CPU as each kernel is concurrently invoked for different parts of the octree, meaning we achieve multicore usage even when using the Serial execution space, with each core working on a separate kernel invocation. Right now, Octo-Tiger can still be compiled without Kokkos and simply uses the old (purely HPX) compute kernel implementations. However, we plan to make Kokkos a mandatory dependency and drop the old, redundant kernel implementations.

\subsection{Octo-Tiger}
To provide context for this work, we briefly introduce Octo-Tiger, its solvers, and its data structure.

Octo-Tiger is used to simulate and study binary star systems and their eventual outcomes.
Stars are modeled in Octo-Tiger as self-gravitating astrophysical fluids.
The Octo-Tiger simulation required implementing two interleaved solvers, a hydro solver and a gravity solver. The hydro solver uses finite volumes to compute the inviscid Navier-Stokes equations. The gravity solver uses a grid-based fast multipole method to compute the (Newtonian) gravitational field generated by the fluid in each time step.

The simulations take too much computational power to use a regular grid. Therefore, we employ adaptive mesh refinement (AMR) to maximize the resolution in the area between the stars, where the mass transfer takes place.
As previously mentioned, we use an adaptive octree as our data structure. Each node in the octree contains a $8 \times 8 \times 8$ sub-grid for computational efficiency. For more astrophysical details, we refer to~\cite{marcello2021octo}.

Put into context with HPX and Kokkos, this means each tree-node (and its sub-grid) is a single HPX component, meaning we can call its functions (for example, the compute kernels) on parent or child tree-nodes without having to worry about which compute node they reside in by using HPX's remote function calls. The (Kokkos) compute kernels each operate on one sub-grid (and potentially its ghost layers if required) at a time, meaning that in each solver iteration, we invoke each compute kernel numerous times (usually once per sub-grid). This gives us numerous parallel kernel calls (tasks) to overlap with the communication and hide associated latencies. This in turn helps with the distributed scalability.

The computational aspects of Octo-Tiger and its scalability were previously studied on Piz Daint~\cite{10.1145/3295500.3356221} and ORNL's Summit~\cite{diehl2021octo}. Recent CPU-related work includes integrating SVE SIMD types for A64FX SIMD~\cite{daiss2022simd} and consequently doing distributed runs on the A64FX machines Stonybrook's Ookami and Riken's Supercomputer\ Fugaku~\cite{10.1145/3585341.3585354}. On the GPU side, we recently worked on dynamic GPU work aggregation \cite{daiss2022aggregation} and did the backend work to integrate HPX with SYCL, potentially targeting Intel GPUs in the future~\cite{10.1145/3585341.3585354}.
While we do not discuss the physics explored by Octo-Tiger in this paper, we note that Astrophysical work with Octo-Tiger include, for example, studies of R Coronae Borealis stars~\cite{staff2018role} and bipolytropic stars~\cite{kadam2018numerical}.

\section{In-house RISC-V test system}
\label{sec:in:house:cluster}
\begin{figure}[tb]
    \centering
    \includegraphics[width=\linewidth]{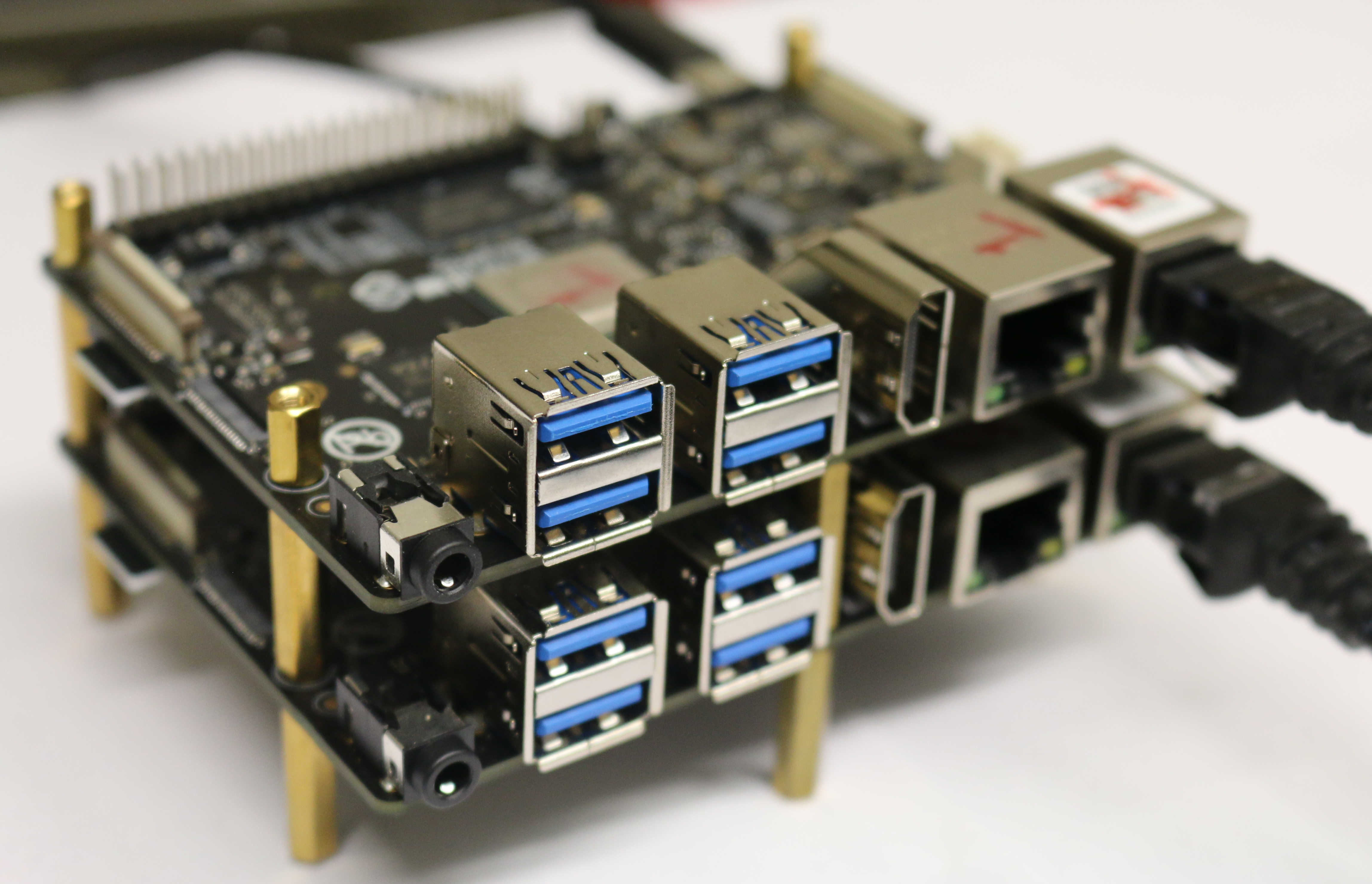}
    \caption{Image of the in-house cluster using two VisionFive2 Open Source RISC-V single board computers with Quad-core StarFive JH7110 64-bit CPU and 8GB LPDDR4 System Memory.}
    \label{fig:inhouse:cluster}
\end{figure}

For this paper, we built an in-house cluster using two VisionFive2 Open Source RISC-V\footnote{\url{https://www.starfivetech.com/en/site/boards}} Single Board Computer (SBC) computers. Each board has Quad-core StarFive JH7110 64-bit CPU and 8 GB LPDDR4 System Memory. The total cost of this test system was \$180. The two boards were connected using the onboard RJ45 Ethernet Connector. Figure~\ref{fig:inhouse:cluster} shows the small in-house cluster. For better cooling, the cluster was placed in an air-conditioned server room. Starfive provides a snapshot of a Debian image\footnote{\url{https://github.com/starfive-tech/Debian}} for this board which works perfectly, but this image does not receive any updates. A warning on the documentation explains that updating the image would break the system.
Although the official image was working perfectly, the integration was challenging. The Slurm workload manager on the Debian was too old, and the FreeIPA client didn't work correctly. Our attempt to upgrade some of the packages to address the issues broke the system.

Fortunately, Ubuntu Linux provides an alternative operating system for VisionFive2\footnote{\url{https://ubuntu.com/download/risc-v}}. The Ubuntu image is the $23.04$ release compiled for RISC-V boards and receives regular updates and the latest packages like other Ubuntu releases. With this image, FreeIPA client worked perfectly with the FreeIPA on RHEL $8.8$, and we could configure the Slurm workload manager. The operating system image was installed on a 64 GB micro SDK with the user home directories and libraries shared over NFS shared file system. As of this writing, the Ubuntu image does not support USB and PCIe on the VisionFive2. The Universal asynchronous receiver-transmitter (UART) serial console was used to configure the network and set up the users, then the rest of the configuration was done over the network. The Star64 and SiFive HiFive Unmatched boards are operated by Tactical Computing Lab and were used to port HPX to RISC-V architecture.

\section{Porting the software stack to RISC-V}
\label{sec:porting:software:stack}

Most of HPX is implemented using ISO C\texttt{++}. However, small portions of the runtime system are implemented using assembly. One portion of HPX that often requires custom assembly is the runtime system's implementation of user-space threads. The HPX context-switching software implementation can optionally utilize \texttt{Boost.Context}\footnote{\url{https://www.boost.org/doc/libs/1_82_0/libs/context/doc/html/index.html}} support or a native independently provided assembly implementation for a targeted ISA. We use \texttt{Boost.Context} in our port of HPX in this paper.

HPX provides software and hardware timing support. The software implementation is portable and utilizes standard ISO C\texttt{++}. Hardware-supported implementations require fewer instructions when compared to software implementations and, as a result, experience performance benefits. 
 
\begin{figure}[tb]
    \centering
    \includegraphics[width=\linewidth,trim={0 1cm 0 1cm},clip]{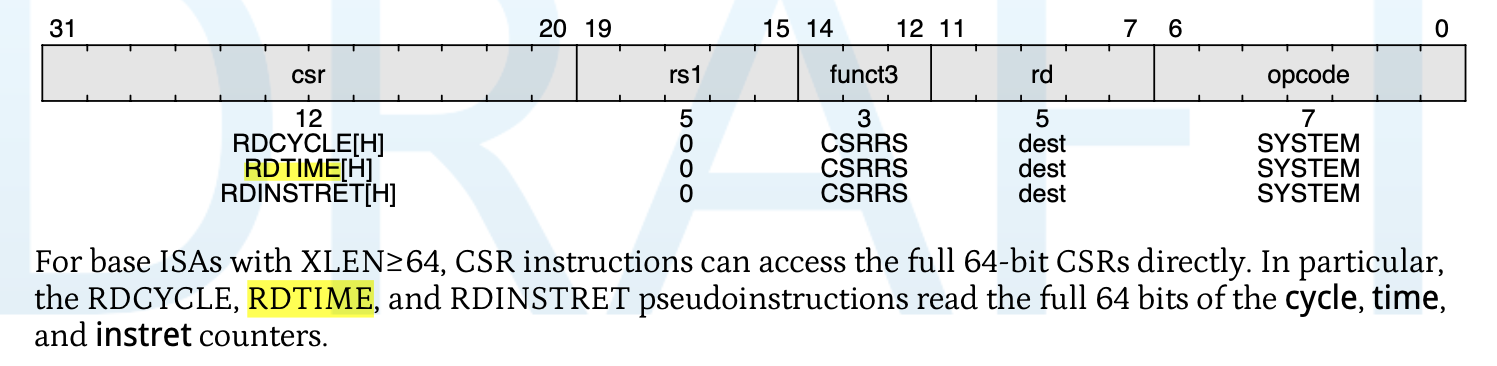}
    \caption{RDTIME instruction from the 2023 Draft Unprivileged ISA. Screenshot from~\cite{riscvisamanual}.}
    \label{fig:rdtime}
\end{figure}

The HPX RISC-V port required making a single source code modification in the HPX timer implementation\footnote{\url{https://github.com/STEllAR-GROUP/hpx/pull/5968}}. Listing~\ref{lst:rdtime} is the RISC-V-specific portion of HPX that adds hardware-supported timers. The RISC-V HPX port implements timing using the RISC-V \lstinline{RDTIME} instruction. \lstinline{RDTIME} is a pseudo-instruction that reads bits from the time Control and Status Register (CSR)~\cite{riscvisamanual}. Figure~\ref{fig:rdtime} is derived from the RISC-V 2023 draft unprivileged ISA and shows the \lstinline{RDTIME} instruction format.

\begin{lstlisting}[language=c++,caption=use of rdtime in HPX,label=lst:rdtime,escapechar=|,float=tbp]
namespace hpx::util::hardware {

    [[nodiscard]]HPX_HOST_DEVICE inline
    std::uint64_t timestamp()
    {
        std::uint64_t val = 0;
        __asm__ __volatile__(
                "rdtime %0;\n"
                : "=r"(val)
                :: );
        return val;
    }
}    // namespace hpx::util::hardware
\end{lstlisting}

The RISC-V port of HPX was initially implemented and tested on a SiFive HiFive Unmatched development board\footnote{\url{https://www.sifive.com/boards/hifive-unmatched}} from 2018 using Ubuntu $22.04.2$. Table~\ref{tab:software} lists the software versions used in this paper. The RISC-V clusters used for this paper include processors made by both the SiFive and StarFive vendors. This paper demonstrates the portability of the HPX RISC-V implementation and the current maturity of RISC-V compiler support.

Porting Kokkos required no changes to the code base, and gcc compiled the library without issues. However, Kokkos's build system CMake files required some minor changes. The RISC-V architecture was not detected, and incorrect compiler flags were added for the architecture and vectorization. A pull request to add RISC-V support providing the correct compiler flags is in preparation\footnote {\url{https://github.com/kokkos/kokkos/issues/6323}}. The RISC-V processors used in this paper do not implement the Vector (V) extension or the SIMD (P) extension in hardware.

As HPX and Kokkos primarily handle the platform-specific details, compiling Octo-Tiger itself for RISC-V was straightforward after these were ported. Our contributions to the HPX and Kokkos toolchains should ease the RISC-V development of other applications.

As a result of this work, HPX, Kokkos, and Octo-Tiger should all be ready to run on future RISC-V-based HPC systems.

\section{Performance measurements}
\label{sec:performance}
In this section, we first compare the performance of HPX's parallel algorithms, sender \& receiver, and asynchronous programming on RISC-V on a SiFive HiFive Unmatched from 2018 having a 4-core U74-MC CPU with the performance on Intel, AMD, and Arm CPUs based on previous work. For more details about the implementations, we refer to~\cite{10.1007/978-3-031-32316-4_3}. Second, we show performance on RISC-V (VisionFive2 Open Source RISC-V) for an astrophysics application, Octo-Tiger, to simulate stellar mergers. Table~\ref{tab:software} offers the software and compiler versions.

\begin{table}[tb]
\caption{Compiler and software versions. The black versions are for the SiFive HiFive Unmatched board and the \textcolor{azure}{versions} for the Pine64\textbackslash VisionFive2 boards. On VisionFive2 gcc $12.2.0$ and OpenMPI $4.1.4$ was used.}
    \label{tab:software}
    \centering
    \begin{tabular}{ccccc}\toprule
     gcc    & HPX & Boost & tcmalloc & hwloc  \\\midrule
      $11.3.0 $ & \textit{d1042a9} & $1.79$/\textcolor{azure}{$1.82$} & $9.9.5$/\textcolor{azure}{$1.11.13$} &  $2.7.0$/\textcolor{azure}{$2.10$}  \\\midrule
      Kokkos & HPX-Kokkos  & cppuddle & jemalloc & Octo-Tiger  \\
      \textit{7a18e97} &  \textit{246b4b8} & 
\textit{c084385} & $5.2.1$ & \textit{aa38039}
      \\\bottomrule
    \end{tabular}
\end{table}

\subsection{HPX Benchmarks}
The C\texttt{++} standard defines approaches for shared memory parallelism: parallel algorithms, asynchronous programming, and senders \& receivers. HPX implements these approaches on top of lightweight threads. For all these approaches, we implemented the Maclaurin series for the natural logarithm $ln$ with the basis $e$, which reads as follows:
\begin{align}
    & ln(1+x) = \notag\\
    &\sum\limits_{n=1}^\infty (-1)^{n+1} \frac{x^n}{n} = x - \frac{x^2}{2} + \frac{x^3}{3} - \ldots, \text{with } \vert x \vert < 1 
    \label{eq:taylor:series}
\end{align}

\begin{figure}[tb]
    \centering

\begin{subfigure}{0.5\textwidth}

\begin{tikzpicture}
\begin{axis}[grid,xlabel={\# cores},ylabel={Flop/s},xmax=10,legend pos=north west,title=\cpp{hpx::async} and \cpp{hpx::future}]

\addplot[black,thick,mark=*, error bars/.cd, y dir=both, y explicit] table [x expr=\thisrowno{0},y expr={100000028581/\thisrowno{2}}, col sep=comma, y error expr={1*(100000028581/\thisrowno{1}-100000028581/\thisrowno{2})}] {ookami_taylor_future_hpx_median.csv};

\addplot[black,thick,mark=triangle*, error bars/.cd, y dir=both, y explicit] table [x expr=\thisrowno{0},y expr={100000028581/\thisrowno{2}}, col sep=comma, y error expr={1*(100000028581/\thisrowno{1}-100000028581/\thisrowno{2})}] {intel_taylor_future_hpx_median.csv};

\addplot[black,thick,mark=square*, error bars/.cd, y dir=both, y explicit] table [x expr=\thisrowno{0},y expr={100000028581/\thisrowno{2}}, col sep=comma, y error expr={1*(100000028581/\thisrowno{1}-100000028581/\thisrowno{2})}] {amd_taylor_future_hpx_median.csv};


\addplot[black,thick,mark=star, error bars/.cd, y dir=both, y explicit] table [x expr=\thisrowno{0},y expr={100000028581/\thisrowno{2}}, col sep=comma, y error expr={1*(100000028581/\thisrowno{1}-100000028581/\thisrowno{2})}] {star64_taylor_future_hpx_median.csv};

\legend{A64FX, Intel,AMD,RISC-V (U74-MC)};
\end{axis}
\end{tikzpicture}
\caption{}
\label{fig:hpx:feautre:performance:async}
\end{subfigure}

\begin{subfigure}{0.5\textwidth}
    \begin{tikzpicture}
\begin{axis}[grid,xlabel={\# cores},ylabel={Flop/s},xmax = 10,legend pos=north west,title=\cpp{hpx::for_each} - \cpp{hpx::execution::par}]

\addplot[black,thick,mark=*, error bars/.cd, y dir=both, y explicit] table [x expr=\thisrowno{0},y expr={100000028581/\thisrowno{2}}, col sep=comma, y error expr={1*(100000028581/\thisrowno{1}-100000028581/\thisrowno{2})}] {ookami_taylor_par_hpx_median.csv};

\addplot[black,thick,mark=triangle*, error bars/.cd, y dir=both, y explicit] table [x expr=\thisrowno{0},y expr={100000028581/\thisrowno{2}}, col sep=comma, y error expr={1*(100000028581/\thisrowno{1}-100000028581/\thisrowno{2})}] {intel_taylor_par_hpx_median.csv};

\addplot[black,thick,mark=square*, error bars/.cd, y dir=both, y explicit] table [x expr=\thisrowno{0},y expr={100000028581/\thisrowno{2}}, col sep=comma, y error expr={1*(100000028581/\thisrowno{1}-100000028581/\thisrowno{2})}] {amd_taylor_par_hpx_median.csv};

\addplot[black,thick,mark=diamond*, error bars/.cd, y dir=both, y explicit] table [x expr=\thisrowno{0},y expr={100000028581/\thisrowno{2}}, col sep=comma, y error expr={1*(100000028581/\thisrowno{1}-100000028581/\thisrowno{2})}] {rv64g_taylor_par_hpx_median.csv};


\legend{A64FX,Intel, AMD,RISC-V (U74-MC)};
\end{axis}
\end{tikzpicture}
\caption{}
\label{fig:hpx:feautre:performance:par}
  \end{subfigure}
    \caption{Performance of asynchronous programming using \cpp{hpx::async} and \cpp{hpx::future} (\protect\subref{fig:hpx:feautre:performance:async}) and HPX's parallel \cpp{hpx::for_each} (\protect\subref{fig:hpx:feautre:performance:par}). The performance measurements for all architectures, except RISC-V, were adapted from~\cite{10.1007/978-3-031-32316-4_3}. The Taylor series for the natural logarithm in Equation~\eqref{eq:taylor:series} for $n=1000000000$ and measured 100000028581 floating point operations using \textit{perf} on a single Intel Core.}
    \label{fig:hpx:feautre:performance}
\end{figure}
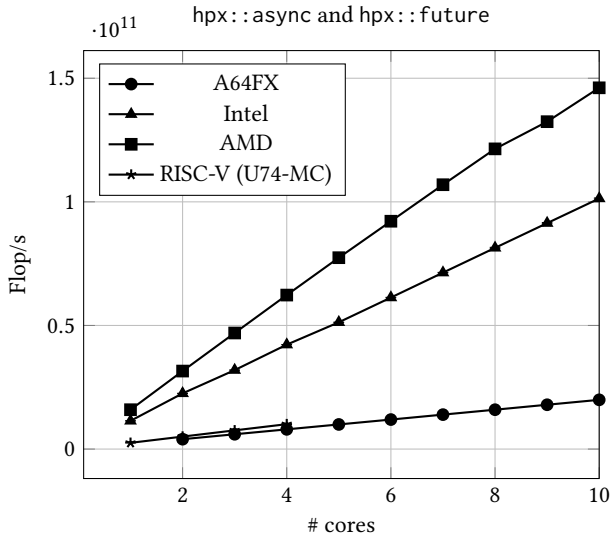
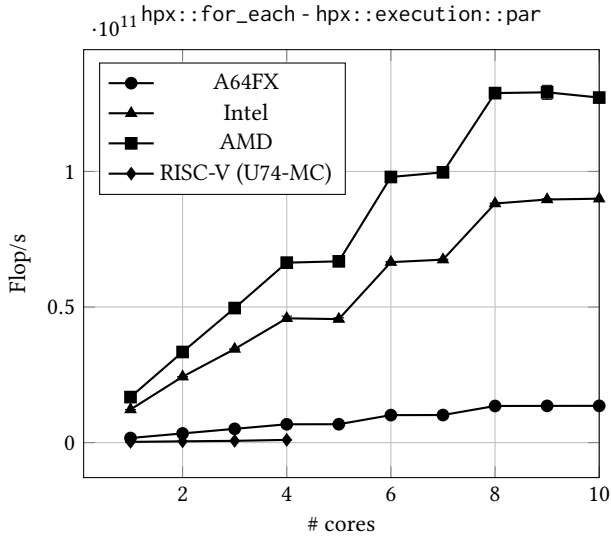

For implementation details, we refer to~\cite{10.1007/978-3-031-32316-4_3} and to the code on GitHub~\cite{patrick_diehl_2023_8067170}, respectively. Figure~\ref{fig:hpx:feautre:performance:async} shows the measured FLOP\textbackslash s using asynchronous programming with \cpp{hpx::async} and \cpp{hpx::future} on all four architectures, namely: Intel Xeon Gold 6140, AMD EPYC 7543, Arm A64FX, and RISC-V SiFive Essential\textregistered U74-MC. We look at the node-level scaling, starting with one core on each platform and then increasing the number of cores used for subsequent runs. Given the limited number of cores on the RISC-V boards, this allows for a fairer comparison to the AMD and Intel machines as we equally limit the utilized cores here. This slightly limits the performance gap between the RISC-V single-board computers and these Intel/AMD production-ready CPUs. 

The code was executed ten times for each core count, and the reported median time was used to calculate the floating point operations per second. We plotted the variance using error bars with minimal time and maximal time. However, the variance is so small that it is only noticable for a few data points. The basis for the floating point operations was measured to be $100000028581$ using a single Intel core using \lstinline[language=bash]{perf}\footnote{\url{https://man7.org/linux/man-pages/man1/perf.1.html}} for $n=1000000000$. Unfortunately, the RISC-V boards do not yet provide hardware counters to measure more accurate floating point numbers. Therefore, we use the value from the Intel core for all other architectures.

The highest performance was measured on AMD, and the second-highest on Intel. The performance of RISC-V was $\approx$ 5 times slower than on A64FX but less compared to AMD and Intel. Figure~\ref{fig:hpx:feautre:performance:par} shows the performance using HPX's parallel algorithm \cpp{hpx::for_each} with the parallel execution policy \cpp{hpx::execution:par}. Here, again, AMD obtained the highest performance, followed by Intel. The performance on RISC-V and A64FX was close but smaller.

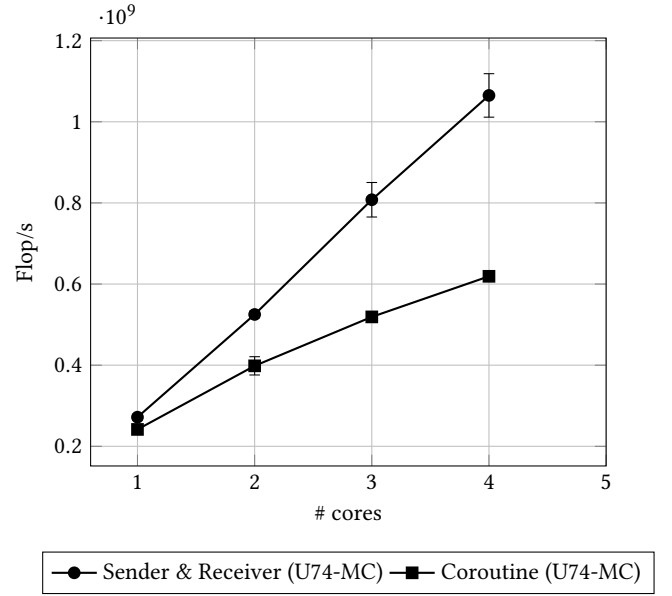
\begin{figure}[tb]

\begin{tikzpicture}
\begin{axis}[grid,xlabel={\# cores},ylabel={Flop/s},xmax=5,legend style={at={(0.5,-0.2)},anchor=north},legend columns=2]



\addplot[black,thick,mark=*, error bars/.cd, y dir=both, y explicit] table [x expr=\thisrowno{0},y expr={100000028581/\thisrowno{2}}, col sep=comma, y error expr={1*(100000028581/\thisrowno{1}-100000028581/\thisrowno{2})}] {rv64g_taylor_sender_receiver_hpx_median.csv};

\addplot[black,thick,mark=square*, error bars/.cd, y dir=both, y explicit] table [x expr=\thisrowno{0},y expr={100000028581/\thisrowno{2}}, col sep=comma, y error expr={1*(100000028581/\thisrowno{1}-100000028581/\thisrowno{2})}] {rv64g_taylor_coroutine_hpx_median.csv};



\legend{Sender \& Receiver (U74-MC),  Coroutine (U74-MC)};
\end{axis}
\end{tikzpicture}
        \caption{Companions of sender \& receiver and future + coroutine parallelism on RISC-V. The Taylor series for the natural logarithm in Equation~\eqref{eq:taylor:series} for $n=1000000000$ and measured $100000028581$ floating point operations using \textit{perf} on a single Intel Core.}
        \label{fig:performance:corroutine}
    \end{figure}

For the last approaches, senders \& receivers and coroutines, a C\texttt{++} compiler supporting the C\texttt{++} 20 standard was needed. Unfortunately, the operating system's default compiler did not support the C\texttt{++} 20 standard on the Intel and the AMD node. The compilers provided as modules did not support the C\texttt{++} 20 standard either. 
We did not compile our customized gcc with the C\texttt{++} 20 standard and deferred these tests to future work. Therefore, Figure~\ref{fig:performance:corroutine} only shows results for RISC-V. The sender \& receiver implementation performed slightly better than the coroutine implementation. The RISC-V hardware supports Fused Multiply Add (FMA) operations; unfortunately, the RISC-V FMA instructions only support the 32-bit floating point ISA. 

To conclude, we have shown that the performance of HPX is around five times less on RISC-V and A64FX. A caveat is that the RISC-V single-board-computer (SBC) is development hardware to port applications to the new CPU architecture. Therefore, the CPU only has four cores. In a previous paper~\cite{10.1007/978-3-031-32316-4_3} using the logarithm benchmark, we plotted more data. However, we capped the data at ten cores to still show the scaling behavior for the RISC-V boards.

\begin{figure}[tb]
    \centering

\begin{subfigure}{0.5\textwidth}

\begin{tikzpicture}
\begin{axis}[grid,xlabel={\# cores},ylabel={$\sfrac{\text{FLOPs(\# cores)}}{\text{Perf}_\text{Peak}\text{(\#cores)}}$},xmax=10,title=\cpp{hpx::async} and \cpp{hpx::future},legend style={at={(0.5,-0.2)},anchor=north},legend columns=2]

\addplot[black,thick,mark=*] table [x expr=\thisrowno{0},y expr={100000028581/\thisrowno{2}/(57.6*\thisrowno{0})/1e9}, col sep=comma] {ookami_taylor_future_hpx_median.csv};

\addplot[black,thick,mark=triangle*] table [x expr=\thisrowno{0},y expr={100000028581/\thisrowno{2}/(73.6*\thisrowno{0})/1e9}, col sep=comma] {intel_taylor_future_hpx_median.csv};

\addplot[black,thick,mark=square*] table [x expr=\thisrowno{0},y expr={100000028581/\thisrowno{2}/(44.8*\thisrowno{0})/1e9}, col sep=comma] {amd_taylor_future_hpx_median.csv};

\addplot[black,thick,mark=diamond*] table [x expr=\thisrowno{0},y expr={100000028581/\thisrowno{2}/(2.4*\thisrowno{0})/1e9}, col sep=comma] {rv64g_taylor_future_hpx_median.csv};


\legend{A64FX, Intel,AMD,RISC-V (U74-MC)};
\end{axis}
\end{tikzpicture}
\caption{}
\label{fig:hpx:feature:performance:async:norm}
\end{subfigure}

\begin{subfigure}{0.5\textwidth}
    \begin{tikzpicture}
\begin{axis}[grid,xlabel={\# cores},ylabel={$\sfrac{\text{FLOPs(\# cores)}}{\text{Perf}_\text{Peak}\text{(\#cores)}}$},xmax = 10,title=\cpp{hpx::for_each} - \cpp{hpx::execution::par},legend style={at={(0.5,-0.2)},anchor=north},legend columns=2]

\addplot[black,thick,mark=*] table [x expr=\thisrowno{0},y expr={100000028581/\thisrowno{2}/(57.6*\thisrowno{0})/1e9}, col sep=comma] {ookami_taylor_par_hpx_median.csv};

\addplot[black,thick,mark=triangle*] table [x expr=\thisrowno{0},y expr={100000028581/\thisrowno{2}/(73.6*\thisrowno{0})/1e9}, col sep=comma] {intel_taylor_par_hpx_median.csv};

\addplot[black,thick,mark=square*] table [x expr=\thisrowno{0},y expr={100000028581/\thisrowno{2}/(44.8*\thisrowno{0})/1e9}, col sep=comma] {amd_taylor_par_hpx_median.csv};

\addplot[black,thick,mark=diamond*] table [x expr=\thisrowno{0},y expr={100000028581/\thisrowno{2}/(2.4*\thisrowno{0})/1e9}, col sep=comma] {rv64g_taylor_par_hpx_median.csv};


\legend{A64FX,Intel, AMD,RISC-V (U74-MC)};
\end{axis}
\end{tikzpicture}
\caption{}
\label{fig:hpx:feautre:performance:par:norm}
  \end{subfigure}
   
    \caption{Normalized performance $\text{Perf}_\text{Norm}\text{(\# cores)}$ by the $\text{Perf}_\text{Peak}\text{(\#cores)}$ for asynchronous programming~(\protect\subref{fig:hpx:feature:performance:async:norm}) and parallel algorithms~\protect(\subref{fig:hpx:feautre:performance:par:norm}), respectively.}
    \label{fig:performance:peak:all}
\end{figure}
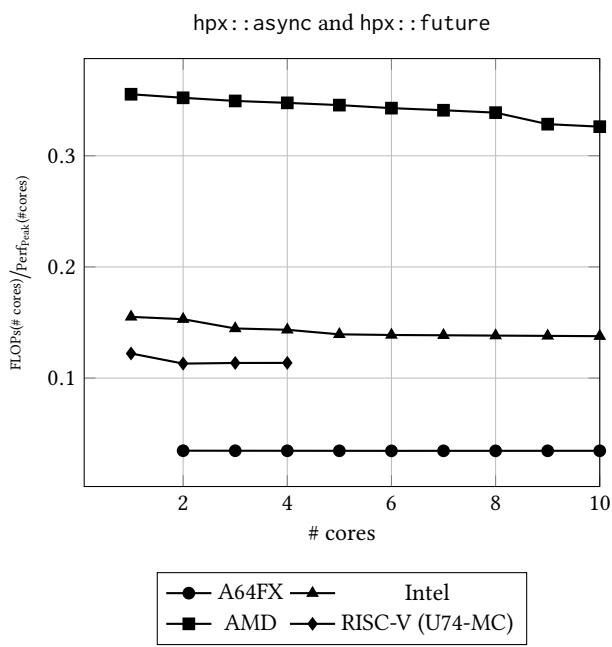
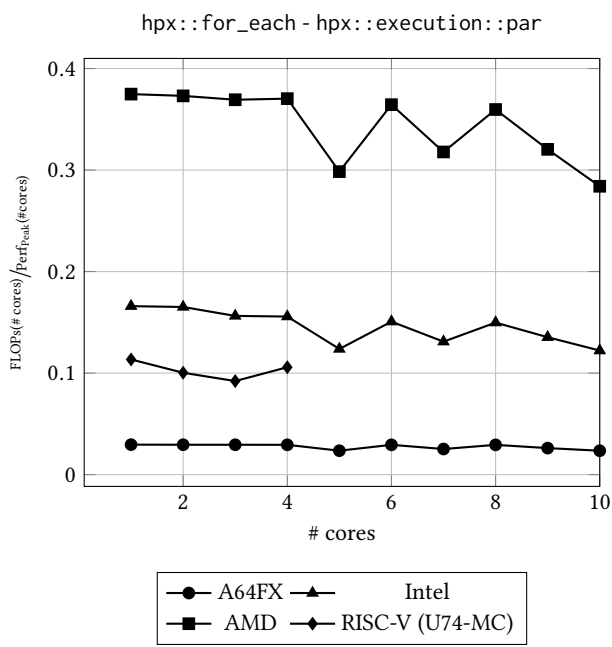

After investigating the FLOP per second, we look into the ratio of floating point operations per second normalized by the theoretical peak performance of each CPU. The theoretical peak performance is given by the clock speed times the vector length times the number of FPU units (\# FPU) and the number of cores (\# cores)
\begin{align}
\label{eq:peak}
    \text{Perf}_\text{Peak}&\text{(\#cores)} = \notag\\
    & 2 \times \text{clock speed} \times \text{vector length} \times \text{\# FPU} \times \text{\# cores} \text{.}
\end{align}
Most CPUs support fused multiplication and addition (FMA), providing a factor of two in Equation~\ref{eq:peak}.
The normalized performance is given by
\begin{align}
    \text{Perf}_\text{Norm}\text{(\# cores)} = \frac{\text{FLOPs(\# cores)}}{\text{Perf}_\text{Peak}\text{(\#cores)}}\text{.}
\end{align}
Table~\ref{tab:peakperformance} lists the clock speed and the vector length obtained by the vendor's data sheets or via \lstinline[language=bash,morekeywords={grep}]{cat /proc/cpuinfo | grep MHz} for all CPUs used in this study. Figure~\ref{fig:hpx:feature:performance:async:norm} shows the normalized performance for asynchronous programming and Figure~\ref{fig:hpx:feautre:performance:par:norm} for parallel algorithms, respectively. For the asynchronous programming and parallel algorithm, we see no effect on auto-vectorization on A64FX. We do not observe a significant effect on Intel and AMD for asynchronous programming. We see some effect on Intel and AMD for the parallel \lstinline[language=c++]{for} loop. Note that both RISC-V CPU do not support vectorization. 

To conclude, the auto-vectorization provided by the GCC compiler does not significantly improve performance. We experienced similar effects for explicit vectorization. For the parallel algorithm, the execution policy \lstinline[language=c++]{hpx::execution::par_unseq} can be used for implicit vectorization~\cite{9651210}. However, this requires the C\texttt{++} 20 standard. For A64FX, support for Scalable Vector Extension (SVE) was shown in~\cite{10029450}. From our experience, explicit vectorization resulted in better performance. 

\begin{table*}[tb]
    \centering
    \caption{Clock speed in GHz, vector length, FPU units per core, fused multiplication and addition (FMA) capability, and number of cores for various CPUs obtained by the vendor specification or via \lstinline[language=bash,morekeywords={grep}]{cat /proc/cpuinfo | grep MHz}. Note the FMS is only available for the 32-bit ISA on RISC-V.}
    \label{tab:peakperformance}
    \begin{tabular}{l|ccccc|c}\toprule
      CPU   & Clock speed [GHz] & Vector length & FPU units per core & FMA & Cores & Peak performance [GFLOP/s] \\\midrule
     ARM A64FX    &  1.8 & 8 & 2 & \checkmark & 48 & 2764.8 \\   
     AMD EPYC 7543 & 2.8 & 4 & 2 & \checkmark & 64 & 2867.2 \\ 
     Intel Xeon Gold 6140 & 2.3 & 8 & 2 & \checkmark & 18 &  1324.8 \\ 
    RISC-V U74-MC(hifiveu) & 1.2 & NA & 1 &  & 4 & 9.6\\\bottomrule
    \end{tabular}
\end{table*}

%
%
%
%
%
%

\subsection{Astrophysics application: Octo-Tiger}
After evaluating HPX, we move to the astrophysics application code Octo-Tiger. We ran a single rotating star with gravity and hydro solvers enabled. We start with node-level scaling using the immense amount of cells fitting in one node from one core up to four cores. Next, we show distributed scaling on two SBCs.

\subsubsection{Node level scaling}
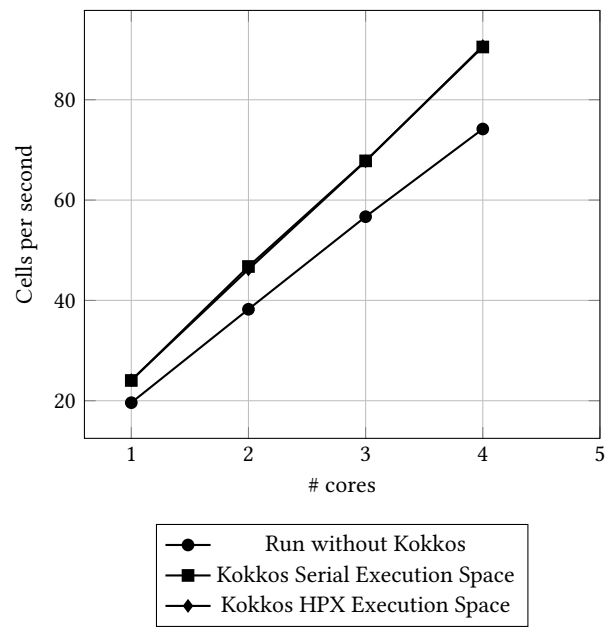
\begin{figure}[tb]

\begin{tikzpicture}
\begin{axis}[grid,xlabel={\# cores},ylabel={Cells per second},xmax=5,legend style={at={(0.5,-0.2)},anchor=north}]

\addplot[black,thick,mark=*] table [x expr=\thisrowno{0},y expr={1184*512/\thisrowno{1}/5}, col sep=comma] {star64-no-kokkos-level5.csv};

\addplot[black,thick,mark=square*] table [x expr=\thisrowno{0},y expr={1184*512/\thisrowno{1}/5}, col sep=comma] {star64-kokkos-level5.csv};

\addplot[black,thick,mark=diamond*] table [x expr=\thisrowno{0},y expr={1184*512/\thisrowno{1}/5}, col sep=comma] {star64-kokkos-hpx-level5};

\legend{Run without Kokkos,Kokkos Serial Execution Space, Kokkos HPX Execution Space};
\end{axis}
\end{tikzpicture}

\caption{Nodel level scaling from one core to all four cores on a single VisionFive2 for rotating star level four, using Octotiger with three different configurations: Once without Kokkos using the old compute kernels, once using the Kokkos Serial execution space and, lastly, once using the Kokkos HPX execution space}
        \label{fig:performance:nodelevel}
    \end{figure}

For the node level scaling, a single rotating star with a level of refinement of four is simulated for five-time steps. The octree has $1184$ leaf nodes with $606208$ cells (as we use 512 cells per sub-grid). Figure~\ref{fig:performance:nodelevel} shows the scaling from a single core to all four cores. First, Octo-Tiger was compiled using native HPX. This test scaled well. Second, Octo-Tiger was compiled using Kokkos and its HPX Execution Space. This test produced similar scaling. Third, we ran the code using the \textit{Kokkos Serial Execution Space}. This test showed some performance improvement over using the HPX Execution space. This is unsurprising since our concurrent kernel launches give us multicore utilization even when using the Serial execution space. Furthermore, the HPX execution space is mostly beneficial in scenarios where we run enough sub-grids for all CPU cores and hence have to divide compute kernel launches into more HPX tasks for full system utilization.

Compared to HPC-grade nodes, we found that the amount of cells computed per second is small (as is expected given the difference in peak performance). The previous section compared the performance of A64FX with less memory usage. With more memory usage, the slow connection to the memory appears to kick in and slows the overall simulation.

\subsubsection{Distributed scaling}
\begin{figure}[tb]
    \centering
\begin{tikzpicture}[scale=0.95, transform shape]
 \begin{axis}[
    xbar=12pt,
    xmin=0,xmax=1100,
    ytick=data,
    enlarge y limits={abs=1cm},
    yticklabels={1-RISC,1-Fugaku,2-RISC-TCP,2-RISC-MPI,2-Fugaku-MPI},
    ytick={1,2,3,4,5},
    bar width = 10pt,
    xlabel= Cells processed per second, 
    ytick align=outside, 
    ytick pos=left,
    major x tick style ={ transparent},
    legend style={at={(0.04,0.96)},anchor=north west, font=\footnotesize, legend cell align=left},
    xmajorgrids=true,
    nodes near coords
        ]    
    \addplot[xbar,fill=cadetgrey!20, area legend] coordinates {
        (91,1.375)
        (168,3.375)
        (140,4.375)
        };
        \addplot[xbar,fill=black!70, area legend] coordinates {
        (778,1.625)
        (1091,4.625)
        };
        
\end{axis}
 
\end{tikzpicture}

    \caption{Distributed scaling for a single node and two nodes using TCP or MPI for communication on RISC-V. Unfortunately, our in-house cluster only had two nodes. For a comparison, runs on a single and two Supercomputer\ Fugaku nodes are shown (each using only four cores out of the 48 available ones for a better comparison).}
    \label{fig:performance:distributed}
\end{figure}
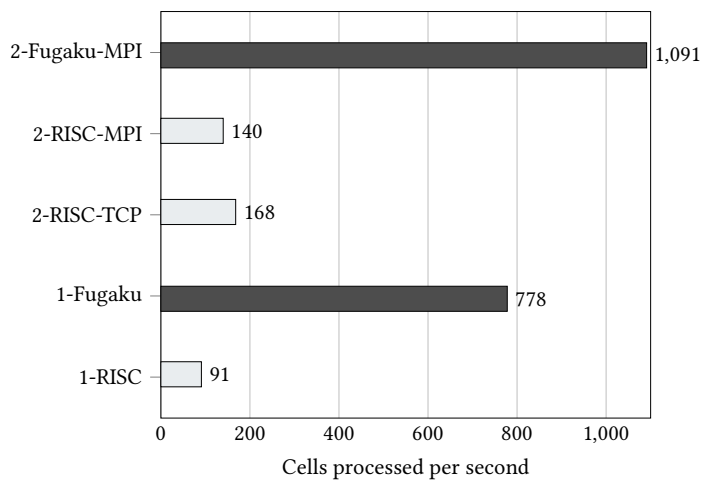

After the node level scaling in the previous experiment, we investigated the scaling on our in-house two single-board computer cluster. We used the rotating star level four and executed it on a single board with all four cores and two boards using all four cores on each. As mentioned in Section~\ref{sec:software:stack}, HPX supports different communication backends. Hence, for our first experiment, we used TCP (Transmission Control Protocol) for communication; and for the second, we used MPI (Message Passing Interface). Figure~\ref{fig:performance:distributed} shows the cells processed per second on a single node and two nodes with different communication backends. The speed-up from a single board to two boards is around $1.85$ for TCP and $1.55$ for MPI. Both communication backends showed some speed-up relative to a single board. However, the difference between TCP and MPI needs further investigation. In addition, we compared the cells processed per second on the Supercomputer\ Fugaku, since, in the previous section, the performance for lower memory intense computations was comparable. For a fair comparison, we used only four cores out of the $48$ cores of the A64FX CPU. On a single node, we observe that the A64FX CPU using the same amount of cores is $\approx 7$ faster.

\section{Energy consumption}
\label{sec:energy}
We compared the RISC-V boards and Supercomputer\ Fugaku for the astrophysics application. On Supercomputer\ Fugaku the power consumption was measured with the PowerAPI~\cite{grant2016standardizing} interface provided by Riken. On the RISC-V boards, no hardware counters for power measurements are present. Here, we attached a power meter to the USB power source and measured the power consumption while running the Linux command \lstinline{stress --cpu 4}\footnote{\url{https://linux.die.net/man/1/stress}} and while running Octo-Tiger with four cores. As a consequence, the power measurements include power for the entire board. This includes the onboard DRAM memory, SSD storage, Ethernet devices, and other components. Power loss occurs at each step of the transition point, from the wall to the power adapter and from the power adapter to the SBC. The ARM PowerAPI isolates the chip's power consumption measurements, which may explain power measurement differences.

\begin{figure}[tb]
    \centering
\begin{tikzpicture}[scale=0.95, transform shape]
 \begin{axis}[
    xbar=12pt,
    xmin=0,xmax=2,
    ytick=data,
    enlarge y limits={abs=1cm},
    yticklabels={1-RISC,1-Fugaku,2-RISC-TCP,2-RISC-MPI,2-Fugaku-MPI},
    ytick={1,2,3,4,5},
    bar width = 10pt,
    xlabel= Watth, 
    ytick align=outside, 
    ytick pos=left,
    major x tick style ={ transparent},
    legend style={at={(0.04,0.96)},anchor=north west, font=\footnotesize, legend cell align=left},
    xmajorgrids=true,
    nodes near coords
        ]    
    \addplot[xbar,fill=cadetgrey!20, area legend] coordinates {
        (71.3/60,1.375)
        (77.1/60,3.375)
        (92.1/60,4.375)
        };
    \addplot[xbar,fill=black!70, area legend] coordinates {
        (55.3/60,1.625)
        (87.31/60,4.625)
        };
\end{axis}
 
\end{tikzpicture}

    \caption{Energy consumption on RISC-V and A64FX CPUs on a single node and two nodes. On Supercomputer\ Fugaku, the power consumption was measured using PowerAPI. Due to missing hardware counters, the power consumption was measured using a power meter on RISC-V.}
    \label{fig:power:distributed}
\end{figure}
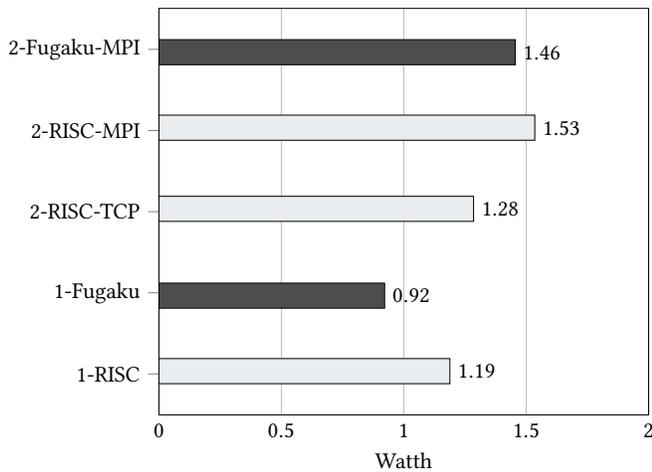

The average power consumption over one minute measured was $3.19$ Watt for \lstinline{stress --cpu 4} and $3.22$ Watt for Octo-Tiger. Figure~\ref{fig:power:distributed} shows the power consumption on both CPU architectures. The power consumption is lower on RISC-V. However, the energy consumption is more significant due to the longer simulation time.



\section{Conclusion}
\label{sec:conclusion}

The RISC-V devices used in this paper are single-board computers (SBC) and development board hardware. The RISC-V devices are manufactured with memory controllers and memory channel counts that are not necessarily competitive with the expectations of HPC-grade hardware consumers. HPX scaled on RISC-V SBCs up to four cores. In comparison, HPX scaling did not yield many performance improvements using the third or fourth core from previous experience~\cite{gupta2020deploying} on Raspberry Pi with an ARM CPU.

Notably, once we got HPX and Kokkos working on RISC-V, we were easily able to obtain a working build of Octo-Tiger, with both the Kokkos compute kernels and its distributed features (via HPX) working. 
While the currently available hardware limits the performance, this paves the way of using Octo-Tiger as a real-world application benchmark for future (potentially HPC-grade) RISC-V hardware. It allows Octo-Tiger users to consider future RISC-V hardware for production runs/simulations.

As for the smaller HPX benchmark (implementing the Maclaurin series),
the runs with small memory usage on the RISC-V boards were around five times slower than on A64FX (when limited to the same number of cores). However, the RISC-V boards were around seven times slower for more memory-intense simulations. For the distributed runs, a more extensive cluster would be of interest. Due to the cost of \$80 per board, this might be expensive for a preliminary portability and evaluation study.

Engineering compromises were made to reduce the single-device price point to move an initial set of devices into the consumer marketplace. The engineering compromises were made to get functional hardware into the marketplace and seed software development efforts such as the development work associated with this paper.

We expect future iterations of RISC-V hardware will provide competitive performance that matches the expectations of the HPC consumer community. For example, the \textit{Milk-V Pioneer}\footnote{\url{https://milkv.io/pioneer}} will have a 64-core \textit{SOPHON SG2042} processor advertised as a desktop machine for development. However, this machine will have 64 cores for larger scaling runs and improved memory and network controllers. It might not be HPC-grade hardware but it will still be a good improvement.

The benchmark for this study makes heavy use of exponentiation. Exponentiation in RISC-V is performed in software as opposed to hardware. Adding hardware support for exponents can reduce the number of floating point operations from approximately ceil((2*e) + 3) down to 4. We believe micro architectural changes, different pipelining techniques, and additional support for out-of-order execution can significantly impact Flops. Hardware counters on RISC-V to measure floating point operations would be beneficial for more accurate estimates.

This paper demonstrates an opportunity for future work that uses memory system benchmarks (GUPS, STREAM, STREAM-Triad, and LINPACK) to grade the relative performance of RISC-V, development board hardware, and HPC-grade devices.

In addition to the hardware development, the development of ISA extensions is ongoing within the RISC-V community.
Some examples that would benefit HPX and other AMTs are one-cycle context switches, extended atomics, hardware support for global address space, and possibly hardware support for thread scheduling (hardware queues).

\begin{acks}
This work was partly funded by DTIC Contract FA8075-14-D-0002/0007 and the Center of Computation \& Technology at Louisiana State University. This research used computational resources of the supercomputer Fugaku provided by the RIKEN Center for Computational Science. The authors would like to thank Stony Brook Research Computing and Cyberinfrastructure, and the Institute for Advanced Computational Science at Stony Brook University for access to the innovative high-performance Ookami computing system, which was made possible by a \$5M National Science Foundation grant (\#1927880).
\end{acks}

\bibliographystyle{ACM-Reference-Format}
\bibliography{software,sample-base}

\appendix

\section{Supplementary materials}
The source code and scripts for benchmarking HPX's feature are available on GitHub\footnote{\url{https://github.com/STEllAR-GROUP/parallelnumericalintegration}}. Octo-Tiger\footnote{\url{https://github.com/STEllAR-GROUP/octotiger}} and HPX\footnote{\url{https://github.com/STEllAR-GROUP/hpx}} are available on GitHub, respectively. The scripts and input files for the distributed runs are available on GitHub\footnote{\url{https://github.com/diehlpkpapers/RISC-V-23}} or Zenodo~\cite{patrick_diehl_2023_8190837}, respectively. The input data is available on Zenodo~\cite{patrick_diehl_2023_8111772}.

\section{Distributes runs without job scheduler}
The cluster was installed without any job scheduler, like Slurm, and we had to provide the configuration for the distributed runs on the command line. For MPI, we could use the \lstinline{--hostfile} option and provide the IP addresses of the nodes within the file. For the TCP runs, Listing~\ref{lst:supervisor} shows the command line options for the supervisor node, and Listing~\ref{lst:delegate} the command line options for the delegate node, respectively. The \textcolor{azure}{blue} IP address is the supervisor, and the \textcolor{amaranth}{red} is the delegate.

\begin{lstlisting}[language=bash,caption=HPX command line option for the supervisor node using TCP,label=lst:supervisor,escapechar=|,float=tb]
./octotiger --config_file=rotating_star.ini 
--max_level=4 --stop_step=5 --theta=0.5 
--multipole_host_kernel_type=KOKKOS  
--monopole_host_kernel_type=KOKKOS 
--hydro_host_kernel_type=KOKKOS 
--hpx:agas=|\textcolor{azure}{10.x.x.160:7910}| 
--hpx:hpx=|\textcolor{azure}{10.x.x.160:7910}| 
--hpx:localities=2 --hpx:threads=4    
\end{lstlisting}

\begin{lstlisting}[language=bash,caption=HPX command line option for the delegate node using TCP,label=lst:delegate,escapechar=|,float=tb] 
octotiger --config_file=rotating_star.ini 
--max_level=1 --stop_step=5 --theta=0.5  
--multipole_host_kernel_type=KOKKOS  
--monopole_host_kernel_type=KOKKOS 
--hydro_host_kernel_type=KOKKOS 
--hpx:agas=|\textcolor{azure}{10.x.x.160:7910}|  
--hpx:hpx=|\textcolor{amaranth}{10.x.x.168:7910}| 
--hpx:worker --hpx:threads=4
\end{lstlisting}

\vspace{-1cm}

\end{document}